\documentclass[aps,pra,showpacs,twocolumn,groupedaddress]{revtex4}
\newcommand \beq{\begin{eqnarray}}
\newcommand \eeq{\end{eqnarray}}
\def\simge{\mathrel{%
       \rlap{\raise 0.511ex \hbox{$>$}}{\lower 0.511ex \hbox{$\sim$}}}}
\def\simle{\mathrel{
       \rlap{\raise 0.511ex \hbox{$<$}}{\lower 0.511ex \hbox{$\sim$}}}}
\usepackage{graphics}
\usepackage[dvips]{graphicx}
\usepackage{graphicx}

\begin{document}

\title{Spin-correlation functions in ultracold paired atomic-fermion
systems: \\ sum rules, self-consistent approximations, and mean fields}
\author{Zhenhua Yu and Gordon Baym}
\affiliation{Department of Physics, University of Illinois,
1110 W. Green Street, Urbana, IL 61801}

\date{\today}

\begin{abstract}

    The spin response functions measured in multi-component fermion gases by
means of rf transitions between hyperfine states are strongly
constrained by the symmetry of the interatomic interactions.  Such
constraints are reflected in the spin f-sum rule that the response
functions must obey.  In particular, only if the effective
interactions are not fully invariant in SU(2) spin space, are the
response functions sensitive to mean field and pairing effects. We
demonstrate, via a self-consistent calculation of the spin-spin
correlation function within the framework of Hartree-Fock-BCS
theory, how one can derive a correlation function explicitly obeying
the f-sum rule.  By contrast, simple one-loop approximations to the
spin response functions do not satisfy the sum rule, except in
special cases.  As we show, the emergence of a second peak at higher
frequency in the rf spectrum, as observed in a recent experiment in
trapped $^6\text{Li}$, can be understood as the contribution from
the paired fermions, with a shift of the peak from the normal
particle response proportional to the square of the BCS pairing gap.


\end{abstract}

\maketitle

\section{Introduction}

    Exploring pairing and superfluidity in ultracold trapped
multicomponent-fermion systems poses considerable experimental and theoretical
challenges \cite{jila_bcs, mit_bcs, duke_capacity, jila_noise, georg}.
Recently, Chin et al.~\cite{chin_rf} have found evidence, by rf excitation,
of a pairing gap in a two-component trapped $^6$Li gas over a range of
coupling strengths.  The experiment, concentrating on the lowest lying
hyperfine states, $|\sigma\rangle = |1\rangle$, $|2\rangle$ and $|3\rangle$,
with $m_F$ = 1/2, -1/2, and -3/2 respectively, measures the long wavelength
spin-spin correlation function, and is analogous to NMR experiments in
superfluid $^3$He \cite{nmr1, nmr2}.  While at high temperatures the rf field
absorption spectrum shows a single peak from unpaired atoms, at sufficiently
low temperature a second higher frequency peak emerges, attributed to the
contribution from BCS paired atoms.  Theoretical calculations at the
``one-loop" level of the spin response \cite{torma,levin,pieri} support
this interpretation.

    In this paper we carry out a fully self-consistent calculation of the
spin-spin correlation function relevant to the rf experiment, at the
Hartree-Fock-BCS level, in order to understand the dependence of the response
on mean field shifts and the pairing gap.  The calculation requires going
beyond the one-loop level, and summing bubbles to all orders, and is valid in
the weakly interacting BCS regime, away from the BEC-BCS crossover -- the
unitarity limit.  An important constraint on the mean field shifts was brought
out by Leggett \cite{leggett} via a sum-rule argument:  For a system with an
interaction that is SU(2)-invariant in spin space, the spins in the
long-wavelength limit simply precess as a whole at the Larmor frequency,
without mean field effects; then the spin-spin correlation function is
dominated by a single pole at the Larmor frequency.  While the effective
interactions between the three lowest hyperfine states of $^6$Li are not
SU(2)-invariant the f-sum rule obeyed by the spin-spin correlation function
still, as we shall show, implies strong constraints on the spin response,
which are taken into account via a self-consistent calculation.

    In order to bring out the physics of a self-consistent approach to the
spin response, we consider a spatially uniform system, and work within the
framework of simple BCS theory on the ``BCS side" of the Feshbach resonance
where the interactions between hyperfine states are attractive.  We assume an
effective Hamiltonian in terms of the three lowest hyperfine states explicitly
involved in the experiments \cite{chin_rf,mit_rf} (we take $\hbar=1$
throughout):
\begin{eqnarray}
    H &=&\int d\mathbf{r} \big\{ \sum_{\sigma=1}^3
  \left(\frac{1}{2m}
  \nabla\psi^\dagger_\sigma(\mathbf{r})\cdot\nabla\psi_\sigma(\mathbf{r})
  +\epsilon^{\sigma}_z\psi^\dagger_\sigma
  (\mathbf{r})\psi_\sigma(\mathbf{r})\right)
  \nonumber\\
  &&+\frac12 \sum_{\sigma,\sigma'=1}^3
    \bar g_{\sigma\sigma'}\psi^\dagger_{\sigma}(\mathbf{r})
  \psi^\dagger_{\sigma'}(\mathbf{r})
  \psi_{\sigma'}(\mathbf{r})\psi_{\sigma}(\mathbf{r})\big\},
 \nonumber\\
 \label{ch}
\end{eqnarray}
where $\psi_\sigma$ is the annihilation operator for state
$|\sigma\rangle$, $\bar g_{\sigma\sigma'}$ is the {\em bare} coupling constant
between states $\sigma$ and $\sigma'$, which we assume to be constant up to a
cutoff $\Lambda$ in momentum space.  Consistent with the underlying symmetry
we assume $\Lambda$ to be the same for all channels, and take
$\Lambda\to\infty$ at the end of calculating physical observables.  The
renormalized coupling constants $g_{\sigma\sigma'}$ are related to those of
the bare theory by
\begin{equation}
  g^{-1}_{\sigma\sigma'}=\bar g^{-1}_{\sigma\sigma'}+m\Lambda/2\pi^2,
  \label{re}
\end{equation}
where, in terms of the s-wave scattering length $a_{\sigma\sigma'}$,
$g_{\sigma\sigma'} = 4\pi a_{\sigma\sigma'}/m$.  In evaluating frequency
shifts in normal states, we implicitly resum particle-particle ladders
involving the bare couplings and generate the renormalized couplings.
However, to treat pairing correlations requires working directly in terms of
the bare $\bar g_{\sigma\sigma'}$ \cite{legg}.

    It is useful to regard the three states $|\sigma\rangle$ as belonging to a
pseudospin (denoted by $Y$) multiplet with the eigenvalues $m_\sigma$ of $Y_z$
equal to 1,0,-1 for $\sigma$ = 1,2,3.  In terms of $m_\sigma$ the Zeeman
splitting of the three levels is
\beq
 \epsilon_Z^\sigma=\epsilon_Z^2 -(\epsilon_Z^3-\epsilon_z^1)m_\sigma/2
    +(\epsilon_Z^3+\epsilon_Z^1 - 2\epsilon_Z^2)m_\sigma^2/2 .
\eeq
The final term in $^6$Li is of order 4\% of the middle term on the BCS side.
The interatomic interactions in the full Hamiltonian for the six $F=1/2$
and $F=3/2$ hyperfine states are invariant under the SU(2) group of spin
rotations generated by the total spin angular momentum $\mathbf{F}$.  The
effective Hamiltonian can be derived from the full Hamiltonian by integrating
out the upper three levels.  However, because the effective interactions
between the lower three levels depend on the non-SU(2) invariant coupling of
the upper states to the magnetic field, the interactions in the effective
Hamiltonian (\ref{ch}) are no longer SU(2) invariant \cite{length, shizhong}.

    In the Chin et al. experiment equal numbers of atoms were loaded into
states $|1\rangle$ and $|2\rangle$ leaving state $|3\rangle$ initially empty;
transitions of atoms from $|2\rangle$ to $|3\rangle$ were subsequently induced
by an rf field.  Finally the residue atoms in $|2\rangle$ were imaged, thus
determining the number of atoms transferred to $|3\rangle$.  The experiment
(for an rf field applied along the $x$ direction) basically measures the
frequency dependence of the imaginary part of the correlation function
$(-i)\int d^3 r\langle T\left(\psi^\dagger_2(\mathbf{r},t)\psi_3(\mathbf{r},t)
\psi^\dagger_3(0,0)\psi_2(0,0)\right)\rangle$ (although in principle atoms can
make transitions from $|2\rangle$ to $|4\rangle$; such a transition, at higher
frequency, is beyond the range studied in the experiment, and is not of
interest presently).  Here $T$ denotes time ordering.  This correlation
function can be written in terms of long-wavelength pseudospin-pseudospin
correlation function, the Fourier transform of
\begin{equation}
    \chi_{xx}(t) = -i\langle T \left(Y_x(t)Y_x(0)\right)\rangle;
 \label{chi}
\end{equation}
here $Y_x = \int d^3r y_x(\mathbf{r})$ is the $x$ component of the total
pseudospin of the system,
\begin{eqnarray}
   y_x(\mathbf{r}) &=&
\frac{1}{\sqrt{2}}\big(\psi^\dagger_1(\mathbf{r})\psi_2(\mathbf{r})
  +\psi^\dagger_2(\mathbf{r})\psi_1(\mathbf{r})\nonumber\\
  &&+\psi^\dagger_2(\mathbf{r})\psi_3(\mathbf{r})+
  \psi^\dagger_3(\mathbf{r})\psi_2(\mathbf{r})\big)
\end{eqnarray}
is the local pseudospin density along the x-axis.  Since the experiment is
done in a many-body state with $N_1=N_2$, the contribution from transitions
between $|1\rangle$ and $|2\rangle$ is zero \cite{SY}.  The Fourier transform
of $\chi_{xx}(t)$ has the spectral representation,
\beq
  \chi_{xx}(\Omega)=\int^\infty_{-\infty}\frac{d\omega}{\pi}
  \frac{{\chi}''_{xx}(\omega)} {\Omega-\omega},
 \label{ft}
\end{eqnarray}
where $\chi''_{xx}(\omega)={\rm Im}\chi_{xx}(\omega-i0^+)$.

    In the next section we discuss the f-sum rule in general, review Leggett's
argument, and illustrate how the sum rule works in simple cases.  Then in
Section III we carry out a systematic calculation, within Hartree-Fock-BCS
theory, of the spin-spin correlation functions, generating them from the
single particle Green's functions.  In addition to fulfilling the f-sum rule,
our results are consistent with the emergence of the second absorption peak
observed in the rf spectrum at low temperature from pairing of fermions.

\section{Sum rules}

    The f-sum rule obeyed by the pseudospin-pseudospin correlation function
arises from the identity,
\beq
 \int^{+\infty}_{-\infty}\frac{d\omega}{\pi}\omega{\chi}''_{xx}(\omega)
                      =\langle [[Y_x,H],Y_x]\rangle.
 \label{sum2}
\eeq
    The need for self-consistency is driven by the fact that the commutator on
the right side eventually depends on the single particle Green's function,
whereas the left side involves the correlation function.  The static
pseudospin susceptibility, $-\chi_{xx}(0)$, is related to
${\chi}''_{xx}(\omega)$ by
\beq
  \chi_{xx}(0)=-\int^\infty_{-\infty}\frac{d\omega}
  {\pi}\frac{{\chi}''_{xx}(\omega)} {\omega}.
 \label{sum1}
\eeq

    Leggett's argument that an SU(2) invariant system gives an rf signal only
at the Larmor frequency is the following:  Let us assume that the
$\bar g_{\sigma\sigma'}$ are all equal, so that the interaction in
Eq.~(\ref{ch}) is SU(2) invariant in pseudospin space; in
addition, let us assume, for the sake of the argument, that the
Zeeman energy is $-\gamma m_\sigma B_z$ ($\gamma$ is the
gyromagnetic ratio of the pseudospin).  Then the right side of
Eq.~(\ref{sum2}) becomes $\gamma B_z\langle Y_z\rangle$, while the
static susceptibility, $-\chi_{xx}(0)$, equals $\langle
Y_z\rangle/\gamma B_z$.  In this case, the spin equations of
motion imply that the response is given by a single frequency (as
essentially found experimentally \cite{nmr1, nmr2}). Thus for
$\omega>0$, we take $\chi''_{xx}(\omega)$ to be proportional to
$\delta(\omega-\omega_0)$.  Combining Eqs.~(\ref{sum2}) and
(\ref{sum1}), we find $\omega_0=\gamma B_z$, the Larmor frequency.
The sum rule implies that neither mean field shifts nor pairing
effects can enter the long wavelength rf spectrum of an SU(2)
invariant system.

    It is instructive to see how the sum rule (\ref{sum2}) functions in
relatively simple cases.  We write the space and time dependent spin density
correlation function as
\begin{eqnarray}
  &D_{xx}(10)\equiv -i\langle T \left(y_x(1) y_x(0)\right)\rangle
  \nonumber\\
   &=\frac12[D_{12}(1)+D_{21}(1)+D_{23}(1)+D_{31}(1)],
\label{dxx}
\end{eqnarray}
where
\begin{eqnarray}
  D_{\beta\alpha}(1)\equiv & -i\langle T\left(
  \psi^\dagger_\alpha(1)\psi_\beta(1)\psi^\dagger_\beta(0)
  \psi_\alpha(0)\right)\rangle,
 \label{D}
\end{eqnarray}
and $\alpha,\beta=1,2,3$.
Here $\psi(1)$, with $1$ standing for $\{\mathbf{r}_1,t_1\}$, is in the
Heisenberg representation, with Hamiltonian $H'=H-\sum_\sigma\mu_\sigma
N_\sigma$.
 Equation~(\ref{dxx}) implies that
$\chi''_{xx}$ is a sum of $\chi''_{\beta\alpha}$, where
\beq
  \chi_{\beta\alpha}(\Omega)\equiv
  \frac{V}{2} D_{\beta\alpha}(\mathbf{q}=0,\Omega+\mu_\alpha-\mu_\beta),
\label{chiD}
\eeq
and $V$ is the system volume.

    As a first example we consider free particles (denoted by superscript
$0$). For $\alpha\neq\beta$,
\begin{eqnarray}
   D^0_{\beta\alpha}(1)=& -iG^0_\alpha(-1)G_\beta^0(1).
\end{eqnarray}
where $G^0_\alpha(1)$, the free single particle Green's function, has
Fourier transform, $G^0_\alpha(\mathbf{k},z)=1/(z-e^\alpha_k)$, with $z$ the
Matsubara frequency and $e^\alpha_k=k^2/2m+\epsilon_Z^\alpha-\mu_\alpha$.
Then,
\beq
 \chi^0_{\beta\alpha}(\Omega)=\frac{1}{2}
 \frac{N_\alpha-N_\beta}{\Omega+\epsilon_Z^\alpha-\epsilon_Z^\beta},
\eeq
from which we see that ${\chi^{0}}''_{\beta\alpha}(\omega)$ has a delta
function peak at $\epsilon_Z^\beta-\epsilon_Z^\alpha$,
as expected for free particles.  This result is manifestly consistent
with Eq.~(\ref{sum2}).

    Next we take interactions into account within the Hartree-Fock
approximation (denoted by $H$) for the single particle Green's function, with
an implicit resummation of ladders to change bare into renormalized coupling
constants.  It is tempting to factorize $D$ as in the free particle case as
\cite{torma,levin,pieri,btrz,OG},
\begin{eqnarray}
  D^{H0}_{\beta\alpha}(1)= -i G^H_\alpha(-1)G_\beta^H(1),
\end{eqnarray}
where $G^H_\alpha(\mathbf{k},z)=1/(z-\zeta^\alpha_k)$, with
\beq
   \zeta^\alpha_k=\frac{k^2}{2m}+\epsilon_Z^\alpha
    +\sum_{\beta(\neq\alpha)}g_{\alpha\beta}n_\beta-\mu_\alpha
\eeq
and $n_\beta$ the density of particles in hyperfine level $\beta$.
Then
\beq
  D^{H0}_{\beta\alpha}(\mathbf{q}=0,\Omega)=\frac{n_\alpha-n_\beta}
 {\Omega+\zeta_0^\alpha-\zeta_0^\beta},
\eeq
and
\beq
\chi_{\beta\alpha}(\Omega)= \hspace{180pt}\nonumber\\
 \frac12
 \frac{N_\alpha-N_\beta}{\Omega+\epsilon_Z^\alpha+
 \sum_{\sigma(\neq\alpha)}g_{\alpha\sigma}n_{\sigma}
 -\epsilon_Z^\beta-\sum_{\sigma'(\neq\beta)}g_{\beta\sigma'}n_{\sigma'}}.
\nonumber\\
\eeq
Consequently
\begin{eqnarray}
\chi''_{\beta\alpha}(\omega)
  &=&-\frac{\pi}{2}(N_\beta-N_\alpha)  \delta(\omega -\Delta E_{\beta\alpha}),
 \label{chih}
\end{eqnarray}
where
\beq
   \Delta E_{\beta\alpha} =
    \epsilon_Z^\beta+\sum_{\sigma'(\neq\beta)}g_{\beta\sigma'}n_{\sigma'}
   -\epsilon_Z^\alpha
   -\sum_{\sigma(\neq\alpha)}g_{\alpha\sigma}n_{\sigma}
\eeq
is the energy difference of the single particle levels $|\alpha\rangle$
and $|\beta\rangle$.  The response function $\chi''_{\beta\alpha}(\omega)$ is
non-zero only at $\omega=\Delta E_{\beta\alpha}$.

    On the other hand, $\chi''_{\beta\alpha}(\omega)$ obeys the sum rule
\begin{eqnarray}
&&\int^{+\infty}_{-\infty}
  \frac{d\omega}{\pi}\omega{\chi}''_{\beta\alpha}(\omega)\nonumber\\
&&      =\frac{V}{2}\int d^3\mathbf r \langle
[[\psi^\dagger_\alpha(\mathbf r)\psi_\beta(\mathbf
r),H],\psi^\dagger_\beta(0)\psi_\alpha(0)]\rangle
 \\
&&  = \frac12(N_\alpha-N_\beta)
  \Big(\Delta E_{\beta\alpha} - g_{\alpha\beta}
  (n_\beta-n_\alpha)\Big).
 \label{sum3}
\end{eqnarray}
where the final line holds for the Hartree-Fock approximation.  The
sum rule (\ref{sum3}) is violated in this case unless
$g_{\alpha\beta}=0$.

    The self-consistent approximation for the correlation function (detailed
in the following section) that maintains the sum rule and corresponds to the
Hartree approximation for the single particle Green's function includes a sum
over bubbles in terms of the renormalized $g$'s:
\beq
  D^H_{\beta\alpha}(q,\Omega)=\frac{
  D^{H0}_{\beta\alpha}(q,\Omega)} {1+g_{\beta\alpha}
  D^{H0}_{\beta\alpha}(q,\Omega)}.
\label{DH}
\eeq
Then with (\ref{DH}),
\begin{eqnarray}
  \chi_{\beta\alpha}{''}(\omega)=\frac{\pi}{2}(N_\alpha-N_\beta)
  \delta\Big(\omega+(\epsilon_Z^\alpha-\epsilon_Z^\beta   \nonumber\\
       +\sum_{\sigma(\neq\alpha)}g_{\alpha\sigma}n_{\sigma}
   -\sum_{\rho(\neq\beta)}g_{\beta\rho}n_\rho+g_{\alpha\beta}
  (n_\beta-n_\alpha))\Big).
  \label{result}
\end{eqnarray}
Note that $\chi_{32}{''}(\omega)$ peaks at
$\omega_H=\epsilon_Z^3-\epsilon_Z^2+(g_{13}-g_{12})n_1$, indicating that the
mean field shift is $(g_{13}-g_{12})n_1$.

    This result agrees with the rf experiment done in a two level $^6{\rm Li}$
system away from the resonance region \cite{mit_two}.  This experiment finds
that no matter whether the atoms in states $|1\rangle$ and $|2\rangle$ are
coherent or incoherent, the rf signal of the transition between $|1\rangle$
and $|2\rangle$ never shows a mean field shift.  As explained in
\cite{mit_two}, in a coherent sample, the internal degrees of freedom of all
the fermions are the same, and thus there is no interaction between them.  In
the incoherent case, the above calculation gives
$\chi_{12}{''}(\omega)=(\pi/2)(N_2-N_1)
\delta(\omega+\epsilon_Z^2-\epsilon_Z^1)$, always peaking at the difference of
the Zeeman energy, and therefore without a mean field contribution.
\cite{hydrogen}

    In an rf experiment using all three lowest hyperfine states, the mean
field shifts appear in $\chi^H_{32}{''}$ as $(g_{13}-g_{12})n_1$.  Since
$g_{\sigma\sigma'}=4\pi a_{\sigma\sigma'}\hbar^2/m$, our result
$(g_{13}-g_{12})n_1$ agrees with Eq.~(1) of Ref.~\cite{mit_rf}.  However, from
$B$ =660 to 900 G (essentially the region between the magnetic fields at which
$a_{13}$ and $a_{12}$ diverge) no obvious deviation of the rf signal from the
difference of the Zeeman energies is observed in the unpaired state
\cite{mit_rf,chin_private}.  The frequency shifts estimated from the result
(\ref{result}) taken literally in this region do not agree with experiment;
one should not, however, trust the Hartree-Fock mean field approximation
around the unitarity limit.  The disappearance of the mean field shifts in the
unitary regime has been attributed to the s-wave scattering process between
any two different species of atoms becoming unitary-limited \cite{chin_rf,
torma}; however, the situation is complicated by the fact that the two
two-particle channels do not become unitarity limited simultaneously.

\section{Self-consistent approximations}

    References \cite{baym} and \cite{baymkad} laid out a general method to
generate correlation functions self-consistently from the single particle
Green's functions.  To generate the correlation function $\chi_{xx}(t)$,
defined in Eq.~(\ref{chi}), we couple the pseudospin to an auxiliary field
$F(\mathbf{r},t)$, analogous to the rf field used in the experiments, via
the probe Hamiltonian :
\beq
  H_{\rm probe}(t)= &-\int d\mathbf{r}
   F(\mathbf{r},t)y_x(\mathbf{r}).
\eeq
The single particle Green's function is governed by the Hamiltonian $H'$,
together with the probe Hamiltonian.  The procedure is to start with an
approximation to the single particle Green's function, and then generate the
four-point correlation function by functional differentiation with respect to
the probe field.  Using this technique we explicitly calculate the
pseudospin-pseudospin correlation functions consistent with the
Hartree-Fock-BCS approximation for the single particle Green's function, in a
three-component interacting fermion system, relevant to the rf experiment on
the BCS side ($g_{\sigma\sigma'}<0$).

    To calculate the right side of Eq.~(\ref{sum2}) directly, we decompose the
Hamiltonian as $H=H_{\rm invar}+H_{\rm var}$, where $H_{\rm invar}$ is
invariant under SU(2) and the remainder
\begin{eqnarray}
  &&H_{\rm var}=\epsilon_Z^2+(\epsilon_Z^3+\epsilon_Z^1-2\epsilon_Z^2)Y_z^2/2
            -(\epsilon_Z^3-\epsilon_Z^1)Y_z/2 \nonumber\\
     &&+(\bar g_{13}-\bar g_{12})\int\psi_3^\dagger\psi_1^\dagger\psi_1\psi_3
     +(\bar g_{23}-\bar g_{12})\int \psi_3^\dagger\psi_2^\dagger\psi_2\psi_3,
\nonumber\\
\end{eqnarray}
is not invariant.  We evaluate the right side of Eq.~(\ref{sum2}),
$\langle [[Y_x,H_{\rm var}],Y_x]\rangle$, term by term in the case that the
states have particle number $N_1 = N_2 =N$ and $N_3=0$, The Zeeman energy in
$H_{\rm var}$ gives $(\epsilon_Z^3-\epsilon_Z^2)N$, and the second term gives
$(\bar g_{12}-\bar g_{13})\int\langle
\psi_2^\dagger\psi_1^\dagger\psi_1\psi_2\rangle$.

    We factorize the correlation function within the Hartree-Fock-BCS theory
for the contact pseudopotential in (\ref{ch}), implicitly resumming ladders to
renormalize the coupling constant in the direct and exchange terms
\cite{dilute}, to write,
\beq
 (\bar g_{12}-\bar g_{13})&\int&\langle
 \psi_2^\dagger\psi_1^\dagger\psi_1\psi_2\rangle \nonumber \\
  &=&(g_{12}-g_{13})\int\langle
 \psi_2^\dagger\psi_2\rangle\langle\psi_1^\dagger\psi_1\rangle
 \nonumber \\
 &&+(\bar g_{12}-\bar g_{13})\int\langle
 \psi_2^\dagger\psi_1^\dagger\rangle\langle\psi_1\psi_2\rangle.
\eeq
Using Eq.~(\ref{re}), we find a contribution from the second term,
$V(g_{12}-g_{13})(n_2n_1+|\Delta|^2/g_{12}g_{13})$, where
$\Delta\equiv\langle\psi_1\psi_2\rangle/\bar g_{12}$, is the BCS pairing gap
between $|1\rangle$ and $|2\rangle$, assumed to be real and positive.  The
last term gives $ (\bar g_{12}-\bar g_{23})\int\int\int\langle
{\psi_2^{\dagger}}'\psi_3' \psi_3^\dagger\psi_2^\dagger\psi_2\psi_3
{\psi_3^{\dagger}}''\psi_2''\rangle =0$; altogether,
\beq
  \int^\infty_{-\infty} \frac{d\omega}{\pi}\omega
  \chi''_{xx}(\omega)    \hspace{144pt}
  \nonumber\\
    =(\epsilon_Z^3-\epsilon_Z^2)N -
  V(g_{12}-g_{13})\left(n_1n_2+\Delta^2/g_{12}g_{13}\right).
 \label{re1}
\eeq
The absence of $g_{23}$ arises from $N_3$ being zero.  Were all
$g_{\sigma\sigma'}$ equal, the right side of Eq.~(\ref{re1}) would reduce to
$(\epsilon^3_Z-\epsilon^2_Z)N$, as expected.  When the interaction is not
SU(2) invariant both mean field shifts and the pairing gap contribute to the
sum rule, allowing the possibility of detecting pairing via the rf absorption
spectrum.

    We turn now to calculating the full Hartree-Fock-BCS pseudospin-pseudospin
correlation function.  For convenience we define the spinor operator \beq \Psi
= \left(\psi_1,\psi_2,\psi_3,
\psi^\dagger_1,\psi^\dagger_2,\psi^\dagger_3\right), \label{Psi} \eeq and
calculate the single particle Green's function \begin{eqnarray}
G_{ab}(1,1')\equiv (-i)\langle T \Psi_a(1)\Psi_b^\dagger(1') \rangle,
\end{eqnarray}
where $1$ denotes $r_1,t_1$, etc., and the subscripts $a$ and $b$ run from
1 to 6 (in the order from left to right in Eq.~(\ref{Psi}); the subscripts 4,
5, and 6 should not be confused with the label for the upper three hyperfine
states), and $\Psi_a(1)$ is in the Heisenberg representation with Hamiltonian
$H''=H'+H_{\rm probe}(t)$.  For $F(\mathbf r,t)=0$ and with BCS pairing
between $|1\rangle$ and $|2\rangle$,
\begin{eqnarray}
  G=\pmatrix{
   G_{11} & 0 & 0 & 0 & G_{15} & 0\cr
   0 & G_{22} & 0 & G_{24} & 0 & 0\cr
   0 & 0 & G_{33} & 0 & 0 & 0 \cr
   0 & G_{42} & 0 & G_{44} &0 & 0\cr
   G_{51} & 0 & 0 & 0 & G_{55} & 0\cr
   0 & 0 & 0 & 0 & 0 & G_{66}}.
   \label{bcsspgf}
\end{eqnarray}

    To obtain a closed equation for $G_{ab}(1,1')$, we factorize the
four-point correlation functions in the equation of motion for $G$ as before,
treating the Hartree-Fock (normal propagator) and BCS (abnormal propagator)
parts differently.  In the dynamical equation for $G_{11}(1,2)$, the term
$\bar g_{12}\langle
\psi^\dagger_2(1)\psi_2(1)\psi_1(1)\psi^\dagger_1(2)\rangle$ is approximated
as $g_{12}\langle\psi^\dagger_2(1)\psi_2(1)\rangle
\langle\psi_1(1)\psi^\dagger_1(2)\rangle$ for the normal part, but $\bar
g_{12}\langle\psi_2(1)\psi_1(1)\rangle
\langle\psi^\dagger_2(1)\psi^\dagger_1(2)\rangle$ for the abnormal part
\cite{legg}.  Since $n_1=n_2$, $\epsilon_Z^1+g_{12}n_2+g_{13}n_3-\mu_2=
\epsilon_Z^2+g_{12}n_1+g_{23}n_3-\mu_2\equiv-\mu_0$, where $\mu_0$ is the free
particle Fermi energy; $\mu_0$ enters into the single particle Green's
function as usual via the dispersion relation
$E_k\equiv\left[(k^2/2m-\mu_0)^2+\Delta^2\right]^{1/2}$ for the paired states.

    The equation of the single particle Green's function in matrix form is
\begin{eqnarray}
 \int d\bar{1}
 \{{G_0}^{-1}(1\bar{1})-F(1)\tau\delta(1-\bar{1}) \nonumber\\
    -\Sigma(1\bar{1})\} G(\bar{1}1') &=\delta(1-1'),
  \label{equationG}
\end{eqnarray}
where the inverse of the free single-particle Green's function is
\begin{eqnarray}
  {G^0}^{-1}_{ab}(11')= \left(i\frac{\partial}{\partial
                  t_1}+\frac{\nabla^2_1}{2m}\pm\mu_a\right)
                    \delta(1-1')\delta_{ab},
\\
\nonumber \\  \nonumber
\end{eqnarray}
with the upper sign for $a$=1,2,3, and the lower for $a$=4,5,6.
The matrix $\tau$ is
\begin{eqnarray}
 \tau=\frac{1}{\sqrt{2}}\left(
 \begin{array}{rrrrrr}
     0 & 1 & 0 & 0& 0& 0\\
     1 & 0 & 1 & 0& 0& 0\\
     0 & 1 & 0 & 0& 0& 0\\
     0 & 0 & 0 & 0& -1& 0\\
     0 & 0 & 0 & -1& 0& -1\\
     0 & 0 & 0 & 0& -1& 0
 \end{array}\right).
\nonumber \\
\end{eqnarray}

\vspace{72pt}
The self energy takes the form
\begin{widetext}
\begin{eqnarray}
-i\Sigma(11')=\hspace{440pt}\nonumber \\
 \begin{small}
 \pmatrix{
     -g_{12}G_{22}-g_{13}G_{33} & g_{12}G_{12} & g_{13}G_{13} & 0
     & -\bar g_{12}G_{15} & -\bar g_{13}G_{16}\cr
     g_{12}G_{21} & -g_{12}G_{11}-g_{23}G_{33} & g_{23}G_{23}
     & -\bar g_{12}G_{24} & 0 & -\bar g_{23}G_{26} \cr
     g_{13}G_{31} & g_{23}G_{32} & -g_{13}G_{11}-g_{23}G_{22} & -\bar g_{13}G_{34}
     & -\bar g_{23}G_{35} & 0\cr
     0 & \bar g_{12}G_{51} & \bar g_{13}G_{61} & g_{12}G_{22}+g_{13}G_{33}
     & -g_{12}G_{21} & -g_{13}G_{31} \cr
     \bar g_{12}G_{42} & 0 & \bar g_{23}G_{62} & -g_{12}G_{12}
     & g_{12}G_{11}+g_{23}G_{33} & -g_{23}G_{32} \cr
     \bar g_{13}G_{43} & \bar g_{23}G_{53} & 0
     & -g_{13}G_{13} & -g_{23}G_{23} & g_{13}G_{11}+g_{23}G_{22}
     },
\end{small}
\nonumber\\
\end{eqnarray}
\end{widetext}
where here $G_{ab}$ denotes $G_{ab}(11^+)\delta(1-1')$ with
$1^+=\{\mathbf{r}_1,t_1+0^+\}$.

    We generate the correlation functions as
\beq
  D_{ab}(12)=-i\sqrt{2}\left(\frac{\delta G_{ab}(11^+)}{\delta
   F(2)}\right)_{F=0};
   \label{defD}
\eeq
(where the factor $\sqrt2$ cancels that from the coupling of
$F(\mathbf{r},t)$ to the atoms via $y_x$) so that from Eq.~(\ref{equationG}),
\begin{eqnarray}
   D(q,\Omega) &=&\frac{\sqrt{2}}{\beta V}\sum_{k,z} G(k,z)
    \left(\tau\right. \nonumber\\ &&+ \frac{\delta
     \Sigma}{\delta B_{\rm rf}}(q,\Omega)\left.\right)G(k-q,z-\Omega).
 \label{correlation}
\end{eqnarray}
Using Eq.~(\ref{bcsspgf}) in (\ref{correlation}), we derive
\beq
  D_{23}=\frac{ D^0_{23}}{1+g_{23} D^0_{23}},
\label{D23} \\
  D_{12}=\frac{ D^0_{12}}{1+g_{12} D^0_{12}},
\label{D12}
\eeq
where
\beq
  D^0_{23}=
  \Pi_{2233}+\bar g_{13}\frac{\Pi_{2433}\Pi_{6651}}{1-\bar g_{13}\Pi_{6611}},
\eeq
and
\beq
  D^0_{12}=\Pi_{1122}-\Pi_{1542};
\eeq
the bubble $\Pi_{abcd}(q,\Omega)$ is given by
\begin{eqnarray}
 \Pi_{abcd}(q,\Omega)=\frac{1}{\beta V}\sum_{k,z}
  G_{ab}(k,z)G_{cd}(k-q,z-\Omega),
\end{eqnarray}
and the summation on $k$ is up to $\Lambda$.  When $\Delta\to 0$,
Eqs.~(\ref{D23}) and (\ref{D12}) reduce to (\ref{DH}), since
$\Pi_{2433}$ and $\Pi_{6511}$ are both proportional to $\Delta$.
Furthermore, when the interaction is SU(2) invariant,
$\chi''_{32}(\omega)$ is proportional to
$\delta(\omega-(\epsilon_z^3-\epsilon_z^2))$.  If only $\bar g_{12}$
is non-zero, the response function $D_{23}$ reduces to the single
loop, $\Pi_{2233}$ (as calculated in Ref.~\cite{pieri}), and in fact
satisfies the f-sum rule (\ref{sum2}).

    To see that the result (\ref{D23}) for the correlation function obeys the
sum rule (\ref{re1}), we expand Eq.~(\ref{D23}) as a power series in
$1/\Omega$ in the limit $\Omega \to \infty$ and compare the coefficients of
$1/\Omega^2$ of both sides.  In addition, with $n_1=n_2$, we find $\int
(d\omega/2\pi)\omega\chi''_{12}(\omega)=0$.

\begin{figure}
\begin{center}\vspace{0cm} \rotatebox{0}{\hspace{-0.cm} \resizebox{6.5cm}{!}
{
\includegraphics{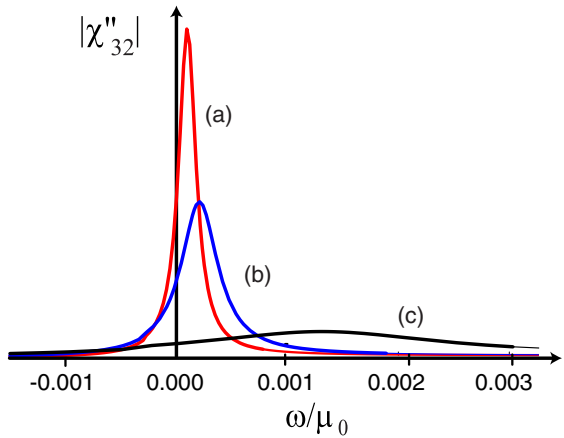}}}
\vspace{1cm} \caption{(Color online). The pseudospin response
function, $|\chi_{32}''|$, vs.  $\omega$ for $k_F a_{12}=-\pi/4$
and $a_{13}=a_{23}=0.1a_{12}$ for three temperatures.  The curves
correspond to a) $T$ = 0.0831, $\Delta$ = 0.0084, b) $T$ = 0.0830,
$\Delta$ = 0.012, and c) $T$ = 0.0820, $\Delta$ = 0.030, all in
units of the free particle Fermi energy in hyperfine states
$|1\rangle$ and $|2\rangle$. } \label{spectrum}
\end{center}
\end{figure}

    Figure~(\ref{spectrum}) shows the paired fermion contribution to
$\chi_{32}''(\omega)$, calculated from Eqs.  (\ref{D23}) and (\ref{chiD}), as
a function of $\omega$, with $g_{\sigma,\sigma'}=4\pi\hbar^2
a_{\sigma,\sigma'}/m$.  This graph corresponds to the $^6$Li experiment done
in a spatially uniform system.  The origin is the response frequency of
unpaired atoms, which is $\omega_{32}^H = \epsilon_z^3 -
\epsilon_z^2+(g_{13}-g_{12})n_1$.  We have not included the normal particle
response in our calculation and do not show this part of the response in the
figure.  The parameters used are $k_F a_{12}=-\pi/4$ and
$a_{13}=a_{23}=0.1a_{12}$, for which, $T_c=0.084\mu_0$.  As the pairing gap,
$\Delta$, grows with decreasing temperature, the most probable frequency,
$\omega_{\rm pair}$, in the response shifts to higher value.  Within the
framework of BCS theory, we can interpret the peak at higher frequency
observed in the rf experiment as the contribution from the paired atoms.

    We now ask how the most probably frequency $\omega_{\rm pair}$ is related
to the pairing gap $\Delta$.  To do this we use the sum rule (\ref{re1}) on
$\chi''_{xx}(\omega)$, written in terms of $\chi''_{32}(\omega)$.  Since
$\chi''_{xx}(\omega)=\chi''_{32}(\omega)+\chi''_{23}(\omega)$ and
$\chi''_{32}(-\omega)=-\chi''_{23}(\omega)$, we have
\begin{equation}
  \int^{\infty}_{-\infty}\frac{d\omega}{\pi}\omega\chi''_{xx}(\omega)
  =2\int^{\infty}_{-\infty}\frac{d\omega}{\pi}\omega\chi''_{32}(\omega).
\label{app1}
\end{equation}
Formally expanding Eq.~(\ref{D23}) as a power series in $1/\Omega$
and comparing the coefficients of $1/\Omega$ on both sides, we
find
\begin{equation}
 \int^{\infty}_{-\infty}\frac{d\omega}{\pi}\chi''_{32}(\omega)=\langle
  Y_z\rangle/2.
\label{app2}
\end{equation}
Then assuming that the rf peak due to pairing is single and narrow (as
found experimentally), we approximate $\chi''_{32}(\omega)$ as $\pi\langle Y_z
\rangle\delta(\omega-\omega_{\rm pair})/2$.  Using Eqs.~(\ref{re1}),
(\ref{app1}) and (\ref{app2}), we finally find
\begin{equation}
 \omega_{\rm pair}-\omega_H=(g_{13}-g_{12})\frac{\Delta^2}{n_0
    g_{12}g_{13}},
 \label{linear}
\end{equation}
where $n_0=n_1=n_2$.  Thus BCS pairing shifts the spectrum away from the
normal particle peak by an amount proportional to $\Delta^2$.

    Equation~(\ref{linear}) enables one to deduce the pairing gap $\Delta$
from the experimental data in the physical case, $g_{13}\neq0$.  However this
result breaks down for the paired states when $g_{13}=0$, a consequence of the
dependence of the sum rule in Eq.~(\ref{re1}) on the cutoff $\Lambda$ of the
bare model (\ref{ch}).  To see this point, we note that the factor
$(g_{13}-g_{12})/g_{12}g_{13}$ that multiplies $\Delta^2$ in Eq.~(\ref{re1})
arises from the combination of the bare coupling constants $1/\bar g_{12}-\bar
g_{13}/\bar g_{12}^2$; using Eq.~(\ref{re}) we can write this combination in
terms of the renormalized coupling constants and the cutoff as
\begin{equation}
  g_{12}^{-1}-\frac{m\Lambda}{2\pi^2}
  -\frac{(g_{12}^{-1}-m\Lambda/2\pi^2)^2}{g_{13}^{-1} -m\Lambda/2\pi^2}.
\end{equation}
Expanding in $1/\Lambda$ in the limit $\Lambda\to\infty$, we see that the
terms linear in $\Lambda$ cancel, leaving the cutoff-independent result,
$g_{13}^{-1}-g_{12}^{-1}$, as in Eqs.~(\ref{re1}) and (\ref{linear}).
However, if only $\bar g_{12}$ is nonzero in this model then we find the
cutoff-dependent result,
\begin{eqnarray}
 \omega_{\rm pair}-\omega_H&=&\frac{\Delta^2}{n_0 \bar g_{12}}
 =\frac{\Delta^2}{n_0}(\frac{1}{g_{12}}-\frac{m\Lambda}{2\pi^2}).
\end{eqnarray}

    Fitting the measured shift in Ref.~\cite{chin_rf}, Fig.~2, to
Eq.~(\ref{linear}), using the values of $a_{12}$ and $a_{13}$ as functions of
magnetic field given in Ref.~{\cite{mit_rf}}, and assuming that
$g_{\sigma\sigma'} =4\pi\hbar^2 a_{\sigma\sigma'}/m$, we find that for Fermi
energy, $E_F=3.6\mu$K, $\Delta/E_F$ = 0.23 at 904G ($k_Fa_{13}=-1.58$,
$k_Fa_{12}=-3.92$), and 0.27 at 875G (or in terms of the Fermi momentum,
$k_Fa_{13}=-1.69$, $k_Fa_{12}=-6.31$).  Similarly for $E_F=1.2\mu$K,
$\Delta/E_F$ = 0.14 at 904G ($k_Fa_{13}=-0.91$, $k_Fa_{12}=-2.26$), and 0.19
at 875G ($k_Fa_{13}=-0.98$, $k_Fa_{12}=-3.64$).  These values are in
qualitative agreement with theoretical expectations \cite{gaps}, although we
expect corrections to the result (\ref{linear}) in the regime where the $k_Fa$
are not small, and in finite trap geometry.

\section{Conclusion}

    As we have seen, the experimental rf result on the BCS side can be
understood by means of a self-consistent calculation of the pseudospin
response within the framework of BCS theory in the manifold of the lowest
three hyperfine states.  The second peak observed at low temperature arises
from pairing between fermions, with the displacement of the peak from the
normal particle peak proportional to the square of the pairing gap $\Delta$.
The shift of the peak vanishes if the interactions within the lowest three
states is SU(2) invariant.  Although the results given here are for the
particular case of the lowest hyperfine states in $^6$Li, the present
calculation can be readily extended to other multiple component fermion
systems, as well as extended to include effects of the finite trap in
realistic experiments.

    We thank Tony Leggett, Shizhong Zhang, and Cheng Chin for valuable
discussions.  This work was supported in part by NSF Grants PHY03-55014 and
PHY05-00914.


\begin{thebibliography}{20}
\bibitem{dilute} An exact self-consistent calculation that starts from the
bare interaction and sums particle-particle ladders to generate the
renormalized interaction includes terms in the correlation functions
involving internal particle loops, as in the last term in Fig. 4c of
Ref. \cite{baymkad}.  However, these terms are of relative order
$k_Fa$, the Fermi momentum times the relevant scattering length, and
in the weak interaction limit can be neglected compared with the
terms we calculate here.
\end{thebibliography}
\end{document}